\documentclass[11pt]{article}
\usepackage[letterpaper,margin=1in]{geometry}
\usepackage[T1]{fontenc}
\usepackage{lmodern,amsmath,amssymb,amsthm,mathrsfs,bm}
\usepackage[hidelinks]{hyperref}
\newcommand{\CalA}{\mathcal A}
\newcommand{\CalC}{\mathcal C}
\newcommand{\CalG}{\mathcal G}
\newcommand{\CalH}{\mathcal H}
\newcommand{\CalL}{\mathcal L}
\newcommand{\BR}{\mathbb R}
\newcommand{\bq}{\mathscr P}
\newcommand{\bS}{\mathbf S}
\newcommand{\bp}{\mathbf p}
\newcommand{\bn}{\mathbf n}
\newcommand{\bnu}{\boldsymbol\nu}
\newcommand{\bu}{\mathbf u}
\newcommand{\vol}{\operatorname{vol}}
\newcommand{\Diff}{\mathsf{Diff}}
\newcommand{\Vect}{\mathsf{Vect}}
\newcommand{\sdtimes}{\mathbin{\ltimes}}
\newcommand{\beq}{\begin{equation}}
\newcommand{\eeq}{\end{equation}}
\numberwithin{equation}{section}
\newtheorem{proposition}{Proposition}[section]
\newtheorem{theorem}[proposition]{Theorem}
\title{Relativistic Hydrodynamics as an Intersection Problem}
\author{Nikita Nekrasov$^1$ and Paul Wiegmann$^2$\\[6pt]
\small $^1$Simons Center for Geometry and Physics and\\
\small Yang Institute for Theoretical Physics, Stony Brook University,\\
\small Stony Brook, NY 11794-3636, USA\\[4pt]
\small $^2$Leinweber Institute for Theoretical Physics,\\
\small Kadanoff Center for Theoretical Physics and Enrico Fermi Institute,\\
\small University of Chicago, Chicago, IL 60637, USA}
\date{}
\begin{document}
\maketitle
\begin{abstract}
Ideal relativistic hydrodynamics admits a covariant description as an
intersection problem in a symplectic space associated with spacetime.
The independent fields are differential forms of degrees $0,1,3,4$.
Spacetime diffeomorphisms and shifts of the momentum one-form act
Hamiltonianly; the zero locus of their moment map is a coisotropic
subvariety encoding the transport equations. The spacetime metric and
equation of state determine a Lagrangian submanifold, whose intersection
with the coisotropic subvariety is precisely the Lichnerowicz--Carter
system. We explain the geometric meaning of the canonical flow parameter,
derive circulation and Ertel invariants directly from the constraints,
and identify energy--momentum conservation with covariance of the
constitutive Lagrangian. The nonrelativistic limit and a relation of
entropy-level helicity to split Calabi--Yau geometry are briefly
indicated.
\end{abstract}

\section{The intersection paradigm}

The equations of an ideal fluid combine two kinds of information:
transport laws and constitutive relations. The former express particle
conservation, entropy advection, and momentum transport; the latter
specify the metric and thermodynamics. Our purpose is to separate them
geometrically, without choosing time or a spatial slice.

For comparison, recall how an ordinary field theory fits this pattern.
Its equations of motion $\delta I/\delta\Phi=0$ follow from an action
$I[\Phi]$, in which the metric enters as a background parameter. The
cotangent bundle to the space of fields $\Phi$, with its canonical
symplectic form, contains two Lagrangian submanifolds: the zero section,
which is canonical, and the graph of $\delta I$, which carries all
dependence on the metric and couplings. Their intersection is the space
of solutions. The equations of an ideal fluid in Eulerian variables are
not naturally of this form: an unconstrained variational formulation
typically introduces auxiliary potentials or particle labels, while an
Eulerian formulation can instead use constrained variations. Here the same
pattern holds with the zero section replaced by a coisotropic subvariety
$\CalC$ and the graph by a Lagrangian submanifold $\CalL$:
\beq\label{eq:paradigm}
\text{fluid flows}\quad\longleftrightarrow\quad
\CalC\cap\CalL_{e,g}\ \subset\ \bq_M.
\eeq
Here $M$ is four-dimensional spacetime. The symplectic space $\bq_M$
and the constraint $\CalC$ do not require a metric; only
$\CalL_{e,g}$ depends on the Lorentzian metric $g$ and the rest-frame
energy density $e$.

This article presents the central construction of \cite{NW} in a
self-contained form. The essential point is that $\CalC$ is a zero
level of a moment map, while $\CalL_{e,g}$ is a graph generated by a
thermodynamic functional. Thus neither the classification of fluid
coadjoint orbits nor a choice of auxiliary particle labels is needed.
We work at the level of smooth fields and formal infinite-dimensional
symplectic geometry; statements about submanifolds and quotients are
understood on smooth regular strata.
A brief closing remark relates entropy-level helicity to the invariant
pairing in a split analogue of Kodaira--Spencer theory.

\section{The symplectic space of spacetime fields}
\label{sec:space}

Let $M$ be an oriented four-manifold. We assume either that $M$ is
closed or that variations, symmetry generators, and boundary conditions
remove the boundary terms in the integrations by parts below. Define
\beq\label{eq:space}
\bq_M=\Omega^0(M)\oplus\Omega^1(M)\oplus\Omega^3(M)\oplus\Omega^4(M)
\ni(\bS,\bp,\bn,\bnu).
\eeq
The fields are initially independent. On the constitutive Lagrangian
they describe an ideal compressible fluid: entropy per particle $\bS$,
momentum per particle $\bp$, particle flux $\bn$, and the four-form
$\bnu=n_gT\,\vol_g$; see \eqref{eq:constitutive}. No barotropic
restriction on the equation of state is imposed.

The canonical symplectic structure is obtained by integrating the wedge
pairing between $\Omega^k(M)$ and $\Omega^{4-k}(M)$:
\beq\label{eq:symplectic}
\Omega=\int_M\delta\bS\wedge\delta\bnu+
                         \delta\bp\wedge\delta\bn.
\eeq
Here $d$ denotes the spacetime exterior derivative and $\delta$ the
exterior derivative on field space. In particular, evaluated on two
variations $a,b$, this means
\beq\label{eq:evaluated}
\Omega(a,b)=\int_M
a_S b_\nu-b_S a_\nu+a_p\wedge b_n-b_p\wedge a_n.
\eeq
The pairing is nondegenerate in the formal sense appropriate to spaces
of smooth fields. In the polarization with coordinates $(\bS,\bp)$,
$\bq_M$ is a cotangent space; the alternative polarization with
coordinates $(\bS,\bn)$ will be useful for the equation of state.

We restrict to the open set where $\bnu$ is nowhere zero. There is then a unique vector field $V$
such that
\beq\label{eq:fluxvelocity}
\iota_V\bnu=\bn.
\eeq
Where $V\neq0$, equivalently where $\bn\neq0$, the kernel of the
three-form $\bn$ is the line spanned by $V$.
Thus $\bn$ determines unparametrized worldlines, whereas $\bnu$
fixes their parametrization. No metric enters this definition.
If $\gamma(\tau)$ is a flow line, its parameter can equivalently be
specified by $\bnu=d\tau\wedge\bn$ along the flow line
(following Carter; see also \cite{Carter,Markakis}).

\section{Two descriptions of the same flow}
\label{sec:flows}

There is a second construction of parametrized curves using
$(\bS,\bp)$. For a path $\gamma:[\tau_0,\tau_1]\to M$ consider
\beq\label{eq:pathaction}
I_{\bS,\bp}[\gamma]=\int_\gamma\bp+
                   \int_{\tau_0}^{\tau_1}\gamma^*\bS\,d\tau.
\eeq
With fixed endpoints, or boundary conditions killing the endpoint
variation, its stationary paths satisfy
\beq\label{eq:hamilton}
\iota_{\dot\gamma}d\bp=d\bS.
\eeq
Indeed, for a path variation $\eta$, the bulk variation is
$\int_{\tau_0}^{\tau_1}\{d\bp(\eta,\dot\gamma)+d\bS(\eta)\}\,d\tau$.
Where $d\bp$ is nondegenerate, these are Hamiltonian trajectories
with Hamiltonian $-\bS$. The constraint formulation also makes sense
when $d\bp$ is degenerate: the vector field is then supplied by
\eqref{eq:fluxvelocity} rather than by inversion of $d\bp$.

Requiring the two constructions of parametrized flow lines to agree,
and imposing particle conservation, defines
\beq\label{eq:constraint}
\CalC=\left\{(\bS,\bp,\bn,\bnu):\quad
\iota_V\bnu=\bn,\quad \iota_Vd\bp=d\bS,\quad d\bn=0\right\}.
\eeq
The first equation is the definition of $V$, so the independent
constraints are the last two equations. Their immediate consequences are
\beq\label{eq:transport}
V(\bS)=0,\qquad L_V\bnu=0,\qquad L_V\bn=0,\qquad
d(\bS\bn)=0.
\eeq
The first follows by contracting $\iota_Vd\bp=d\bS$ with $V$;
the others follow from Cartan's formula and $d\bn=0$.
Thus the entropy flux $\bS\bn$ is conserved, as is the particle flux.
These statements involve no equation of state.

\medskip
\noindent{\bf Canonical parametrization.}
The parameter in \eqref{eq:pathaction} is part of the geometry.
The shift $\bp\mapsto\bp+df$ changes the functional only by an
endpoint term. The shift $\bp\mapsto\bp+\bS\,df$ is accompanied by
$\tau\mapsto\tau-f(\gamma(\tau))$, and hence changes the velocity to
\beq\label{eq:reparametrization}
V\longmapsto(1-V(f))^{-1}V,
\eeq
where the denominator is nonzero. These transformations preserve the
unparametrized curves, while changing their clocks. This distinction
will explain why the natural circulation law---the extension of the
Kelvin theorem to baroclinic flows (cf.~\cite{Carter,Markakis})---uses
$V$ rather than the unit four-velocity.

Define
\beq\label{eq:zeta}
\zeta=\bS+\iota_V\bp.
\eeq
It is the Lagrangian of \eqref{eq:pathaction} evaluated on a flow line,
$I_{\bS,\bp}[\gamma]=\int_{\tau_0}^{\tau_1}\zeta(\gamma(\tau))\,d\tau$. On the
constitutive Lagrangian of Sec.~\ref{sec:lagrangian} it becomes
$\zeta=-\mu/T$, with $\mu=w-T\bS$ the chemical potential.
On $\CalC$, the momentum satisfies
\beq\label{eq:liep}
L_V\bp=d\zeta.
\eeq
In particular, the two-form $d\bp$ is preserved by the canonical flow.

\section{The coisotropic constraint as a moment-map level}
\label{sec:moment}

The coisotropic nature of \eqref{eq:constraint} follows from symmetry.
The group
\beq\label{eq:group}
\CalG=\Diff(M)\sdtimes C^\infty(M)
\eeq
acts on $\bq_M$ by pullback of all fields and by
\beq\label{eq:shift}
(\bS,\bp,\bn,\bnu)\longmapsto
(\bS,\bp+df,\bn,\bnu).
\eeq
Both actions preserve $\Omega$. We use the Hamiltonian convention
$\iota_{X_H}\Omega=-\delta H$.

For $\epsilon\in\Vect(M)$ and $f\in C^\infty(M)$, Hamiltonians
generating the infinitesimal pullback and shift actions are
\beq\label{eq:hamiltonians}
\begin{split}
H_\epsilon&=-\int_M\left(\bnu\,L_\epsilon\bS+
                              (L_\epsilon\bp)\wedge\bn\right),\\
H_f&=\int_M f\,d\bn.
\end{split}
\eeq
For example, varying $H_f$ gives
$\delta H_f=-\int_M df\wedge\delta\bn$, which shows that $H_f$
generates \eqref{eq:shift}. Invariance of the integral under
diffeomorphisms similarly verifies the first line.

\begin{proposition}\label{prop:moment}
The action of $\CalG$ on $\bq_M$, by pullback and by the shift
\eqref{eq:shift}, has an equivariant moment map with components
\beq\label{eq:moment}
\mu_{C^\infty}=d\bn,\qquad
\mu_{\Diff}=(\iota_Vd\bp-d\bS)\otimes\bnu+
                                \bp\otimes d\bn.
\eeq
Consequently $\CalC=\mu_{\CalG}^{-1}(0)$, where
$\mu_{\CalG}=(\mu_{C^\infty},\mu_{\Diff})$.
\end{proposition}

\begin{proof}
The Abelian component follows from $H_f$. For the diffeomorphism
component, Cartan's formula and integration by parts give
\beq\label{eq:momentcalculation}
\begin{split}
H_\epsilon
&=-\int_M\left(\epsilon(\bS)\bnu+
     (\iota_\epsilon d\bp)\wedge\bn+
     d(\iota_\epsilon\bp)\wedge\bn\right)\\
&=\int_M\bigl((\iota_Vd\bp)(\epsilon)-d\bS(\epsilon)\bigr)\bnu
                     +\bp(\epsilon)d\bn.
\end{split}
\eeq
We used $\bn=\iota_V\bnu$.
This is the pairing of $\epsilon$ with the tensor-valued four-form
$\mu_{\Diff}$ in \eqref{eq:moment}. The Hamiltonians are the canonical
ones for a cotangent lift and a momentum translation, so the moment
map is equivariant; no central term arises under the assumed boundary
conditions. Setting the Abelian component to zero first leaves
$\iota_Vd\bp=d\bS$, proving the assertion.
\end{proof}

An equivalent expression avoids dividing by $\bnu$. Let $\bn^\vee$
denote the volume-form-valued vector field corresponding to $\bn$
under $\Omega^3(M)\simeq TM\otimes\Omega^4(M)$. Then
\beq\label{eq:momentdensity}
\mu_{\Diff}=\iota_{\bn^\vee}d\bp+
                  \bp\otimes d\bn-d\bS\otimes\bnu.
\eeq
This expression is polynomial in the fields and their first derivatives.

\medskip
\noindent{\bf Why coisotropic?}
For an equivariant moment map the Poisson bracket of two component
Hamiltonians is again a component Hamiltonian, with the appropriate
Lie bracket of generators. The constraints are therefore first class:
their Poisson brackets vanish on $\mu_{\CalG}^{-1}(0)$. On regular strata this
is precisely the coisotropic condition
$(T\CalC)^{\perp_\Omega}\subset T\CalC$. The characteristic directions
are generated by the $\CalG$ action. One may form the symplectic quotient
$\mu_{\CalG}^{-1}(0)/\CalG$ where it exists, but the fluid equations themselves
are formulated on $\CalC$ before any quotient is taken.

There is a second realization of the same semidirect-product symmetry,
obtained by replacing \eqref{eq:shift} with
\beq\label{eq:secondshift}
(\bS,\bp,\bn,\bnu)\longmapsto
(\bS,\bp+\bS\,df,\bn,\bnu-df\wedge\bn).
\eeq
It is symplectic and has Abelian moment map $d(\bS\bn)$.
The diffeomorphism moment map is unchanged, and
\beq\label{eq:entropyidentity}
\iota_V\mu_{\Diff}=\zeta\,d\bn-d(\bS\bn).
\eeq
Thus particle conservation implies entropy conservation on
$\mu_{\Diff}=0$; conversely entropy conservation implies particle
conservation wherever $\zeta\ne0$.
Consequently the two realizations of
the Abelian factor, the momentum shift \eqref{eq:shift}, which leaves
the parametrization of flow lines unchanged, and the shift
\eqref{eq:secondshift}, which changes their clock, cut out the same
coisotropic subvariety $\CalC$ wherever $\zeta\ne0$,
that is, on the constitutive Lagrangian, wherever $\mu\neq0$.
The particle-conservation formulation remains valid also where the
chemical potential vanishes.

\section{The constitutive Lagrangian and relativistic Euler equations}
\label{sec:lagrangian}

Now equip $M$ with a Lorentzian metric $g$ of signature $(-,+,+,+)$.
We use the Hodge convention $\alpha\wedge\star\beta=
\langle\alpha,\beta\rangle_g\vol_g$. On the timelike-current sector
write
\beq\label{eq:density}
\bn=n_g\,\iota_{\bu}\vol_g,\qquad
g(\bu,\bu)=-1,\qquad
n_g=\sqrt{\frac{\bn\wedge\star_g\bn}{\vol_g}}>0.
\eeq
Here $\bu$ is the unit four-velocity; for a perfect fluid the Landau
and Eckart frames coincide. Notice the difference with $V$; see
\eqref{eq:physicalvelocity}.
Choose an energy density $e=e(n_g,\bS)$ and define
\beq\label{eq:thermodynamics}
w=\left(\frac{\partial e}{\partial n_g}\right)_{\bS},\qquad
T=\frac1{n_g}\left(\frac{\partial e}{\partial\bS}\right)_{n_g},\qquad
P=n_gw-e.
\eeq
We assume $w,T>0$. The thermodynamic identity is
\beq\label{eq:firstlaw}
de=w\,dn_g+n_gT\,d\bS,\qquad
dP=n_g\,dw-n_gT\,d\bS.
\eeq
The function $e$ and the metric are constitutive data, not part of the
definition of $\CalC$.

Consider the functional
\beq\label{eq:action}
\CalA_{e,g}(\bS,\bn)=-\int_M e(n_g,\bS)\,\vol_g.
\eeq
Its differential defines a submanifold of $\bq_M$ by
\beq\label{eq:constitutive}
\CalL_{e,g}=\left\{\quad
\bp=\frac{w}{n_g}\star_g\bn,\qquad
\bnu=n_gT\,\vol_g\quad\right\}.
\eeq

\begin{proposition}\label{prop:lagrangian}
The submanifold \eqref{eq:constitutive} is Lagrangian for
\eqref{eq:symplectic}, with generating functional
\eqref{eq:action} in the $(\bS,\bn)$ polarization.
\end{proposition}

\begin{proof}
For fixed $g$, varying $n_g$ gives
$n_g\delta n_g\,\vol_g=\delta\bn\wedge\star_g\bn$.
Consequently
\beq\label{eq:variation}
\delta\CalA_{e,g}
=\int_M\frac{w}{n_g}\star_g\bn\wedge\delta\bn
                  -n_gT\vol_g\,\delta\bS
=\int_M\bp\wedge\delta\bn-\bnu\,\delta\bS.
\eeq
The field-space exterior derivative of the last one-form is $\Omega$.
Its restriction to the graph therefore vanishes. Since the graph has
the independent coordinates $(\bS,\bn)$, it is maximal isotropic.
\end{proof}

\begin{theorem}\label{thm:intersection}
On the physical sector \eqref{eq:density}, smooth ideal relativistic
fluid configurations with equation of state $e$ are precisely the
points of $\CalC\cap\CalL_{e,g}$.
\end{theorem}

\begin{proof}
With our Hodge convention, $\star_g\bn=n_g\bu^\flat$.
Equations \eqref{eq:fluxvelocity} and \eqref{eq:constitutive} become
\beq\label{eq:physicalvelocity}
\bp=w\bu^\flat,\qquad V=\bu/T.
\eeq
The remaining constraints are exactly
\beq\label{eq:CL}
d(n_g\iota_{\bu}\vol_g)=0,\qquad
\iota_{\bu}d(w\bu^\flat)=T\,d\bS,\qquad
\bu(\bS)=0.
\eeq
The last equation follows from the second.
These are the covariant
Lichnerowicz--Carter equations \cite{L,Carter,Markakis}.
Conversely a solution of \eqref{eq:CL} defines $\bn,\bp,\bnu$
by \eqref{eq:density} and \eqref{eq:constitutive}, and hence lies
in the stated intersection.
\end{proof}

For completeness, the equivalence with the stress-tensor formulation
is immediate in components. Put $D=\bu^\mu\nabla_\mu$,
$a_\nu=D u_\nu$, and
$\Delta_\nu{}^\mu=\delta_\nu{}^\mu+u_\nu u^\mu$.
The momentum equation in \eqref{eq:CL} reads
\beq\label{eq:acceleration}
w a_\nu+\Delta_\nu{}^\mu\nabla_\mu w=T\nabla_\nu\bS.
\eeq
Using \eqref{eq:firstlaw} and $D\bS=0$, this becomes
$(e+P)a_\nu+\Delta_\nu{}^\mu\nabla_\mu P=0$.
Together with particle conservation it gives
\beq\label{eq:stress}
\nabla_\mu T^{\mu\nu}=0,\qquad
T^{\mu\nu}=(e+P)u^\mu u^\nu+Pg^{\mu\nu}.
\eeq
The component along $\bu$ is
$De+(e+P)\nabla_\mu u^\mu=0$, a consequence of the first law,
$D\bS=0$, and $Dn_g+n_g\nabla_\mu u^\mu=0$.

\medskip
\noindent{\bf Pressure polarization.}
Where the density Legendre transform is regular, the same Lagrangian
has coordinates $(\bS,\bp)$ and generating functional
\beq\label{eq:pressure}
\CalH_{P,g}(\bS,\bp)=-\int_M P(p_g,\bS)\,\vol_g,
\qquad p_g=\sqrt{-g^{-1}(\bp,\bp)}.
\eeq
Indeed, on \eqref{eq:constitutive}, $p_g=w$,
$P_w=n_g$, and $P_{\bS}=-n_gT$. This functional is obtained from
\eqref{eq:action} by the partial Legendre transform
\beq\label{eq:legendre}
\CalH_{P,g}=\int_M\bp\wedge\bn-\CalA_{e,g}.
\eeq
The sign follows from
$\bp\wedge\bn=-n_gw\vol_g$. The different generating functions
therefore describe the same constitutive submanifold, rather than
different fluid equations.

\medskip
\noindent{\bf Covariance as conservation.}
Metric variations of \eqref{eq:action} define
\beq\label{eq:metricvariation}
T_{\mu\nu}=-\frac{2}{\vol_g}\frac{\delta\CalA_{e,g}}{\delta g^{\mu\nu}}.
\eeq
At fixed $\bn$, varying $n_g^2$ and $\vol_g$ yields
the perfect-fluid tensor \eqref{eq:stress}.
Diffeomorphism invariance of $\CalA_{e,g}$, under simultaneous
transformation of the fields and the metric, together with
\eqref{eq:variation}, gives on $\CalL_{e,g}$
\beq\label{eq:covariance}
H_\epsilon=\frac12\int_M T_{\mu\nu}\,L_\epsilon g^{\mu\nu}\,\vol_g
=\int_M\epsilon_\nu\nabla_\mu T^{\mu\nu}\,\vol_g ,
\eeq
that is,
\beq\label{eq:momentstress}
\mu_{\Diff}\big|_{\CalL_{e,g}}=(\nabla_\mu T^{\mu}{}_{\nu})\,dx^\nu\otimes\vol_g .
\eeq
This holds on $\CalL_{e,g}$ alone, before $\CalC$ is imposed. On
$\CalC\cap\CalL_{e,g}$ it gives energy--momentum conservation;
conversely, $d\bn=0$ and $\nabla_\mu T^{\mu\nu}=0$ imply
$\mu_{C^\infty}=\mu_{\Diff}=0$. Thus
the dynamical conservation law is the diffeomorphism moment map
restricted to the constitutive Lagrangian.

\medskip
\noindent{\bf Nonrelativistic limit.}
We briefly comment on the nonrelativistic limit. On a spatial slice
$B^3$, write $\rho=m\,\bn|_{B^3}$ for the mass density, with $m$ the
particle mass, $s=\bS\rho/m$ for the entropy density, and
$\Pi=c^{-1}\bp|_{B^3}\otimes\rho/m$ for the momentum density; the
constraints become coadjoint transport for
$\Diff(B^3)\sdtimes C^\infty(B^3,\BR^2)$ \cite{Novikov,Holm,NW}.
With $g=-c^2dt^2+h$ and the rest energy $mc^2n_g$ separated from $e$,
one has $\Pi\to v^\flat\otimes\rho$, with $v$ the Newtonian velocity.
As $c\to\infty$ the constitutive Lagrangian is generated by the
Newtonian energy: the integral over $B^3$ of the kinetic energy
density $|\Pi|_h^2/2\rho$ plus the internal energy density
$\varepsilon(\rho,s)$, densities being taken with respect to $\vol_h$.
Its intersection with the transport constraint gives the usual
compressible Euler equations, with metric and thermodynamics still
confined to the Lagrangian data.

\section{Circulation, Ertel current, and entropy-level helicity}
\label{sec:invariants}

These consequences of the equations are already visible on $\CalC$.
By \eqref{eq:liep}, a closed contour $C_\tau$ transported by $V$
satisfies
\beq\label{eq:kelvin}
\frac{d}{d\tau}\oint_{C_\tau}\bp=\oint_{C_\tau}d\zeta=0.
\eeq
This is the circulation law in the canonical fluid frame
\cite{Carter,Salmon}. The normalization matters: for a contour
transported by $\bu=TV$, Cartan's formula instead gives
$L_{\bu}\bp=T\,d\bS+d(\iota_{\bu}\bp)$, and the circulation
derivative is $\oint T\,d\bS$, which need not vanish.

The Ertel current and its scalar coefficient are
\beq\label{eq:ertel}
J_E=d\bS\wedge d\bp=Q\bn.
\eeq
To see why $Q$ exists, note that $dJ_E=0$ and
\beq\label{eq:ertelcontraction}
\iota_VJ_E=V(\bS)d\bp-d\bS\wedge\iota_Vd\bp=0.
\eeq
At a point where $V\ne0$, the three-forms annihilated by $V$
form a one-dimensional space, spanned by $\bn$.
Both $J_E$ and $\bn$ are invariant under $V$, so
\beq\label{eq:erteladvection}
V(Q)=0.
\eeq
Where $\bp\wedge\bn\ne0$, one may equivalently write
$Q=(\bp\wedge J_E)/(\bp\wedge\bn)$.
For any smooth function $F$, the current
$F(\bS,Q)\bn$ is closed; its integral on homologous transverse
three-surfaces is therefore conserved under the assumed boundary
conditions. This is the covariant form of potential-vorticity
conservation \cite{Katz}.

The helicity three-form is
\beq\label{eq:helicity}
h=\bp\wedge d\bp.
\eeq
For a regular entropy level $N_s=\bS^{-1}(s)$, the integrals
\beq\label{eq:entropylevel}
H_s=\int_{N_s}\bp\wedge d\bp
\eeq
are invariant under the two momentum shifts
$\bp\mapsto\bp+df$ and $\bp\mapsto\bp+\bS\,df$, provided
$N_s$ is closed or the boundary term vanishes: restricted to $N_s$,
their change is an exact form.

Moreover $V$ is tangent to $N_s$, and
$\iota_V(d\bp|_{N_s})=d\bS|_{N_s}=0$. To identify the associated
volume, locally write $\bnu=d\bS\wedge\eta$ near a regular level.
The pullback $\eta_s=\eta|_{N_s}$ is a well-defined, nowhere-zero
three-form, independent of the choice of $\eta$. Since
$\bn=-d\bS\wedge\iota_V\eta$, equation \eqref{eq:ertel} gives
\beq\label{eq:entropyvolume}
d\bp|_{N_s}=\iota_V\sigma_s,\qquad \sigma_s=-Q\,\eta_s.
\eeq
Where $Q\neq0$, equivalently where the restricted vorticity is
nonzero, $\sigma_s$ is a volume form. It is uniquely determined by
\eqref{eq:entropyvolume} and $V$, and $L_V\sigma_s=0$: indeed,
$d(\iota_V\sigma_s)=d(d\bp|_{N_s})=0$. Thus $V|_{N_s}$ is an
exact divergence-free field for this solution-dependent volume, with
potential $\bp|_{N_s}$, and $H_s$ is its helicity whenever the integral
exists. Under the global hypotheses of Arnold's helicity theorem---for
example, on a closed oriented $N_s$ with its complete volume-preserving
flow---this also has the interpretation of an asymptotic Hopf invariant
\cite{ArnoldKhesin}. The construction applies to entropy levels on
which $Q$ is nowhere zero, or to their nondegenerate regions with
appropriate boundary or decay conditions. On each such region the
canonical flow preserves a naturally determined volume, although the
underlying relativistic fluid remains compressible.
These entropy-level
helicity integrals are natural invariants of the covariant constraint;
one should not identify them with a conserved spatial helicity for a
general non-isentropic compressible fluid.

\section{Split Calabi--Yau geometry and further applications}
\label{sec:bridges}

The split counterpart of Calabi--Yau complex-structure geometry has
complementary real rank-three distributions; their integrability is
an additional differential condition, not a consequence of a stable
three-form alone \cite{Hitchin}. On a local product $B_+\times B_-$
with volume form $\vol_+$ on $B_+$, the formal split analogue of the
volume-preserving Kodaira--Spencer integrability equation is
$d_-\mathsf A+\frac12[\mathsf A,\mathsf A]=0$: here $d_-$ differentiates
on $B_-$, and $\mathsf A$ is a one-form on $B_-$ valued in
divergence-free vector fields on $B_+$; the bracket combines exterior
multiplication with their Lie bracket \cite{BCOV}. On the exact
divergence-free algebra the invariant pairing is
\beq\label{eq:helicitypairing}
\langle v,w\rangle_+=\int_{B_+}\alpha_v\wedge d\alpha_w,
\qquad d\alpha_v=\iota_v\vol_+,
\eeq
assuming $B_+$ is closed, or boundary and decay conditions make the
integral and integrations by parts well defined. This pairing defines
the corresponding Chern--Simons functional, and its quadratic form is
the helicity \cite{ArnoldKhesin}. Section~\ref{sec:invariants} supplies a
fluid realization of this algebra and pairing: with $B_+=N_s$ and
$\vol_+=\sigma_s$ on a nondegenerate entropy level,
$\alpha_V=\bp|_{N_s}$ and $H_s=\langle V,V\rangle_+$. This is a
connection at the level of the invariant pairing; it does not identify
the full Kodaira--Spencer field with a compressible fluid solution.

\medskip
\noindent{\bf Other systems.}
The same pattern appears in several systems discussed in \cite{NW}.
For self-dual fields, middle-degree forms on a Lorentzian
$(4l+2)$-dimensional spacetime carry the symplectic form
${\int\alpha\wedge\beta}$; closed forms constitute a coisotropic
subspace and self-dual forms a metric-dependent Lagrangian one,
whose intersection is the space of closed self-dual forms. The cases
$l=0,1,2$ include the one-form field strength of a chiral boson on the
string worldsheet, the three-form field strength of the M5-brane
tensor field, and the self-dual five-form field strength of type IIB
supergravity. Charged fluids are described either by letting the
constitutive Lagrangian depend on background gauge fields, or by
enlarging the symplectic space to include the electromagnetic field,
which gives the Euler--Maxwell system. Generating functionals
depending on derivatives of $\bn$ and $\bS$ define larger classes
of Lagrangian submanifolds, with the transport constraint unchanged;
their detailed analysis lies beyond the scope of this article.

Anomalous hydrodynamics provides another direction: anomaly-induced
transport and modified current-conservation laws require corresponding
extensions of the constitutive data and the constraint geometry. A way to incorporate the anomaly of the underlying
microscopic degrees of freedom is to add the magnetic term
\beq
 \frac{k}{2}\int_M d\bp\wedge\delta\bp\wedge\delta\bp
\label{eq:magterm}
\eeq
to the symplectic form \eqref{eq:symplectic}.
Here $k$ is a constant. The correction \eqref{eq:magterm} is closed on field space and invariant
under diffeomorphisms and the shift \eqref{eq:shift}; the Abelian moment
map becomes $d\bn-\tfrac{k}{2}\,d\bp\wedge d\bp$, so particle
conservation is replaced by the anomalous law
$d\bn=\tfrac{k}{2}\,d\bp\wedge d\bp$; see \cite{NW} for details
and the corresponding modification of the Lichnerowicz--Carter equations. 

In a different direction, 
we are exploring extensions of the present framework that incorporate viscosity.

\medskip
\noindent{\bf Acknowledgments.}
We thank A.~Abanov, A.~Cappelli, Ya.~Eliashberg, D.~Freed,
B.~Khesin, M.~Kontsevich, R.~Suszek, D.~Sullivan and M.~Zabzine for discussions.
N.~N. was supported by the Simons
Collaboration ``Probabilistic Paths to QFT'', and by the Knut and Alice
Wallenberg Foundation. N. N. thanks Uppsala University and IHES for their hospitality.
P.~W. thanks BIMSA and YMSC at Tsinghua University, where this work
was done, for their hospitality.

\end{document}